\documentclass[aps,pra,twocolumn,superscriptaddress,showpacs,10pt]{revtex4-1}
\usepackage{hyperref, enumerate}
\usepackage{soul}
\usepackage{graphicx}
\usepackage{amsmath}
\usepackage{amssymb}
\usepackage{verbatim}
\usepackage{color}
\usepackage{cancel}

\usepackage{amsthm}

 \theoremstyle{definition}

 \theoremstyle{remark}

\def\be{\begin{equation}}
\def\beq{\begin{eqnarray}}
\def\ee{\end{equation}}
\def\eeq{\end{eqnarray}}
\def\eqref#1{(\ref{#1})}

\begin{document}
\author{Arthur Jaffe}
\affiliation{Harvard University, Cambridge, Massachusetts 02138, USA}
 
\author{Fabio L. Pedrocchi}
\affiliation{Department of Physics, University of Basel, Basel, Switzerland}
\affiliation{Harvard University, Cambridge, Massachusetts 02138, USA}

\title{Topological Order and Reflection Positivity}

\begin{abstract}
The interplay between the two fundamental concepts of topological order and reflection positivity allows one to characterize the ground states of certain many-body Hamiltonians. We define topological order in an appropriate fashion and show that certain operators have positive expectation value in all ground states. We apply our method to vortex loops in a model relevant to topological quantum memories. 
\end{abstract}

\pacs{03.65.-w, 05.30.Fk, 05.30.Pr}

\maketitle

\emph{Introduction.}--- Topologically ordered systems have attracted much attention, since they represent promising candidates for the realization of a fault-tolerant quantum computing architecture. Intuitively, topological order can be understood as the property that local perturbations cannot cause transitions between the different  degenerate ground states of certain many-body Hamiltonians; such transitions require global perturbations. Therefore the subspace of topologically ordered ground states seems to be a good place to store and process quantum information \cite{KitaevToric,BHM,DKLP,NO}.

Another fundamental concept that we use is reflection positivity. This  notion originally arose in the theory of random fields, as the property that justifies \emph{inverse} Wick rotation from random fields to quantum fields \cite{OS}, both at  zero and at positive temperature \cite{HoeghKrohn}.  This property has also played an important role in the analysis of phase transitions and ground states in statistical mechanical systems \cite{GJS,FSS,Lieb1994}.  

In this work, we show that these two fundamental concepts allow one to characterize the ground states of certain reflection-symmetric Hamiltonians. In particular, we show that the expectation value of certain local operators are positive in all ground states. In certain examples this means that the ground states are vortex-free. 

Understanding the ground-state properties of topologically ordered systems is not only interesting from a fundamental point of view, but is also relevant to topological quantum computation: the ground states encode the logical-qubit states.

We apply our method to a Hamiltonian introduced in \cite{THD} which is a quartic polynomial in Majoranas (Majorana operators) defined on a planar lattice. We focus on this interaction because, in lowest-order perturbation theory, it is equivalent to the toric code model, the archetypical model for a topologically-ordered quantum memory~\cite{KitaevToric}. We show in this example that with our choice of the signs of the coupling constants, all the ground states are free of vortices.

We first review the concept of reflection positivity. We then introduce the concept of $W$-topological order that is a special case of the more general topological order defined in \cite{NO,BHM}. With these two concepts in hand, we show how topological order and reflection positivity allow one to characterize the properties of the degenerate ground states. In particular, we show that the expectation value of certain operators is non-negative. Finally, we show that when the Hamiltonian in \cite{THD} is reflection-symmetric, no vortices are present in the ground states. 

\emph{Reflection Positivity.}--- Consider a lattice $\Lambda$ which is divided in two parts $\Lambda_{\pm}$ mapped into each other by reflection in a plane $\Pi$. We represent this reflection as an anti-unitary operator $\vartheta$ on the Hilbert space $\mathcal{H}$ of our model. To each vertex of the lattice, we associate one or more operators $O_{i}$, and the reflection maps them into
\begin{equation}\label{eq:reflection}
\vartheta(O_{i})=\vartheta\,O_{i}\,\vartheta^{-1}=O_{\vartheta i}^{\dagger}\,,
\end{equation}
where site $\vartheta i$ is the reflection of site $i$.

Let $\mathfrak{A}$ denote the set of operators that are sums of products of those $O_{i}$'s, where $i$ is in $\Lambda_{-}$. The reflection-positivity property for the pair $H$ and $\mathfrak{A}$ means: for every $A\in\mathfrak{A}$ and every $0\leqslant\beta$, 
\begin{equation}\label{eq:RP}
0\leqslant \text{Tr}(A\,\vartheta(A)\,e^{-\beta H})\,.
\end{equation}
We use the notation $W_{A}=A\,\vartheta(A)$. 

\emph{$W$-Topological Order.}--- Let us denote the ground-state subspace of $H$ to be $\mathcal{P}$. We also use the symbol $\mathcal{P}$ for the orthogonal projection onto the ground-state subspace. One says that $\mathcal{P}$ has $W$-topological order if $\mathcal{P}W\mathcal{P}$ is a scalar multiple of $\mathcal{P}$. This definition is a specialization of general topological order defined in \cite{NO,BHM}. In other words, the operator $W$ cannot cause transitions between different ground states.\color{black}

\emph{Topological Order ensures Positivity.}--- Consider a Hamiltonian $H$ and an operator $A\in\mathfrak{A}$ with the reflection positivity property (\ref{eq:RP}). Assume that the ground-state subspace has $W_{A}$-topological order. Then we infer positivity 
\begin{equation}
0\leqslant\langle\Omega, W_{A}\Omega\rangle\,,
\end{equation}
for any ground state $\Omega$ of $H$.

\emph{Explanation.}--- Suppose $H$ has $N$ orthonormal ground states denoted by $\Omega^{\mu}$, with $\mu=1,\ldots,N$. Normalize the ground-state energy of $H$ to be zero. Assuming (\ref{eq:RP}) and taking the $\beta\rightarrow\infty$ limit leads to
\begin{equation}\label{eq:pos_ground}
0\leqslant \sum_{\mu=1}^{N}\langle \Omega^{\mu}, W_{A}\,\Omega^{\mu}\rangle. 
\end{equation}
We infer from (\ref{eq:pos_ground}) that the expectation value of $W_{A}$ is non-negative in at least one of the ground states. Since the ground states are $W_{A}$-topologically ordered,
\begin{equation}
\langle\Omega^{\mu},W_{A}\,\Omega^{\mu}\rangle=\alpha \langle\Omega^{\mu},\Omega^{\mu}\rangle=\alpha\,,
\end{equation}
with $\alpha$ a constant, independent of $\mu$. As the expectation value must be non-negative for some $\mu$, we conclude that $0\leqslant\alpha$, and
\begin{equation}\label{eq:positivity}
0\leqslant\langle \Omega^{\mu}, W_{A}\,\Omega^{\mu}\rangle\,\qquad\text{for}\quad\mu=1,\ldots,N\,.
\end{equation}

\emph{Vortex Loops.}---In certain situations, the operator $W_{A}$ equals a vortex loop.
A loop $C$ of length $2l$ is an ordered sequence $\{i_{1},i_{2},\ldots,i_{2l},i_{1}\}$ of nearest-neighbor sites in $\Lambda$. Associated to each loop $C$, we define a vortex loop $W(C)$ as a product of $O_{i}$'s,
\begin{equation}\label{eq:WOi}
W(C)=O_{i_{1}}O_{i_{2}}\cdots O_{i_{2l}}\,.
\end{equation}
Since we want $W(C)$ to have the form $A\vartheta(A)$, we choose $A=O_{i_{1}}\cdots O_{i_{l}}$, so by (\ref{eq:reflection}), 
\begin{equation}
W(C)=O_{i_{1}}O_{i_{2}}\cdots O_{i_{l}}O_{\vartheta i_{1}}^{\dagger}O_{\vartheta i_{2}}^{\dagger}\cdots O_{\vartheta i_{l}}^{\dagger}\,.
\end{equation}

In the following example, we consider the case where each of the operators $O_{i}$ is a Majorana. Then $c_{i}=c_{i}^{\dagger}=c_{i}^{-1}$ and $\{c_{i},c_{j}\}=2\delta_{ij}$. In this case, we introduce a phase $i^{l}$ and redefine $W(C)$ in place of (\ref{eq:WOi}) as
\begin{equation}\label{eq:WC}
W(C)=i^{l}c_{i_{1}}c_{i_{2}}\cdots c_{i_{l}}c_{\vartheta i_{1}}c_{\vartheta i_{2}}\cdots c_{\vartheta i_{l}}\,.
\end{equation}
 Thus, we have   $W(C)=W(C)^{\dagger}=W(C)^{-1}$, so the vortex loop has eigenvalues $\pm 1$, and its expectation value in a unit vector lies between $-1$ and $+1$.

We say that the loop $C$ is vortex-free when the expectation value of $W(C)$ is $+1$ and vortex-full when the expectation value is $-1$. In the intermediate cases, we say that the loop is partially free and partially full, according to the sign of the expectation value of $W(C)$.

In the special case that $W(C)$ commutes with $H$, and therefore $W(C)$ is conserved, it is possible to choose an orthonormal basis of ground states of $H$ for which $W(C)=\pm1$. Then the loop $C$ is either vortex-free or vortex-full in each of these ground states.

\begin{figure}[h!]
\centering\includegraphics[width=0.8\columnwidth]{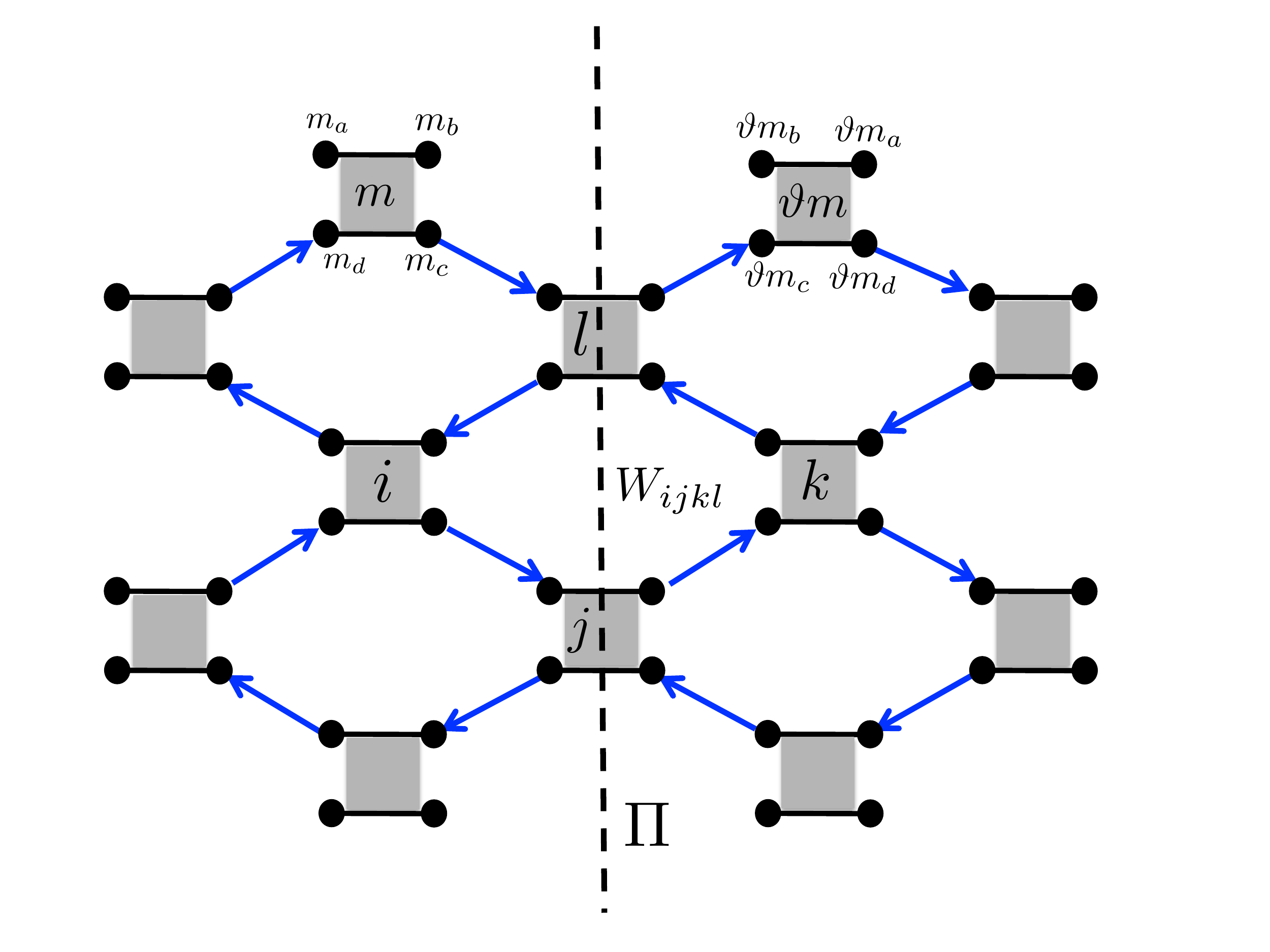}
 \caption{In the lattice, we depict square islands by gray squares and Majoranas by black dots. Bonds $(jk)$ which connect nearest-neighbor square islands are indicated by directed bonds between nearest-neighbor Majoranas. The orientation of the bonds define circuits around eight-site polygons that lie between the square islands.}
  \label{fig:figure_1}
\end{figure}

\emph{An example of topological quantum memory.}---As an example, we consider the Hamiltonian proposed in \cite{THD}, describing interactions between Majoranas localized on the vertices of a planar lattice.
The model studied in \cite{THD} has a Hamiltonian of the form
\begin{equation}\label{eq:Hamiltonian}
H=\sum_{j}H_{0,j}+\lambda\sum_{j<k}V_{(jk)}\,.
\end{equation}
Here $j$ labels square islands of the lattice, see Fig.~\ref{fig:figure_1}, and the Hamiltonian $H_{0,j}$ is a product of four independent Majoranas $c_{j_{a}}$, $c_{j_{b}}$, $c_{j_{c}}$, and $c_{j_{d}}$ of the form,
\begin{equation}
H_{0,j}=- c_{j_{a}}c_{j_{b}}c_{j_{c}}c_{j_{d}}\,.
\end{equation}
The constant $\lambda$ in (\ref{eq:Hamiltonian}) is dimensionless, and  $V_{(jk)}=ic_{j}c_{k}$ are quadratic interactions between Majoranas. 
The authors show in \cite{THD} that for small values of the parameter $\lambda$ in the Hamiltonian, the model possesses $W$-topological order for local operators $W$. 

We illustrate the planar-lattice configuration in Fig.~\ref{fig:figure_1}. Each pair of nearest-neighbor islands $j$ and $k$ defines a directed bond $(jk)$ that characterizes the coupling of the neighboring islands, and which determines $V_{(jk)}$. Each island $j$ consists of a square with independent Majoranas $c_{j}^{a}$, $c_{j}^{b}$, $c_{j}^{c}$, and $c_{j}^{d}$ which we place on the four corners of the square island as specified in Fig.~\ref{fig:figure_1} (in particular $a$, $b$ lie on the top of the square and $c$, $d$ lie on the bottom of the square). 
For four nearest-neighbor squares labeled $i$, $j$, $k$, $l$ on the lattice, define the loop $C_{ijkl}=\{i,j,k,l\}$ and the vortex operator
\begin{equation}
W(C_{ijkl})=c_{i_{c}}c_{j_{a}}c_{j_{b}}c_{k_{d}}c_{k_{a}}c_{l_{c}}c_{l_{d}}c_{i_{b}}\,.
\end{equation}
The vortex operators have the form (\ref{eq:WC}) and thus eigenvalues $\pm1$.

\emph{Majoranas and Reflection Positivity.}---The vortex $W(C_{ijkl})$ that is bisected by the plane $\Pi$, as illustrated in Fig.~\ref{fig:figure_1}, can be written in the form 
\begin{equation}\label{eq:WA}
W(C_{ijkl})=W_{A}=A\,\vartheta(A)\,,
\end{equation}
with $A=c_{l_{d}}c_{i_{b}}c_{i_{c}}c_{j_{a}}$.  With our choice of $V_{(jk)}$ the Hamiltonian $H$ in (\ref{eq:Hamiltonian}) is reflection-symmetric, 
\begin{equation}
\vartheta(H)=H\,.
\end{equation}
In \cite{JP}, we have studied a class of reflection-symmetric Majorana Hamiltonians that includes the Hamiltonian (\ref{eq:Hamiltonian}), and we demonstrated that reflection positivity holds.

As a consequence, the expectation value of $W(C_{ijkl})$ in the thermal state at inverse temperature $\beta$ is positive. Furthermore, in its topological phase, the loop $C_{ijkl}=\{i,j,k,l\}$ is vortex free in all ground states.  When the ground state is non-degenerate, (\ref{eq:RP}) implies that loop $C_{ijkl}$ is vortex-free in the ground state.  Finally, if the lattice is periodic, any loop $C_{ijkl}$ can be used in this argument. Therefore, every ground state  is free of each elementary vortex. 

\emph{Conclusion.}---This work is based on general principles. Therefore the methods described here might be useful in the study of other many-body, topologically-ordered Hamiltonians.

\emph{Acknowledgements.}---We thank Zohar Nussinov and Diego Rainis for useful comments on an earlier version of this manuscript. This work was supported by the Swiss NSF and NCCR QSIT.

\end{document}